\documentclass[sigconf]{acmart}

\usepackage{booktabs} 
\usepackage{balance}
\usepackage{url}

\setcopyright{none}

\renewcommand\footnotetextcopyrightpermission[1]{} 
\begin{document}
\copyrightyear{2019} 
\acmYear{2019} 
\setcopyright{acmlicensed}
\acmConference[ASPDAC '19]{24th Asia and South Pacific Design Automation Conference}{January 21--24, 2019}{Tokyo, Japan}
\acmBooktitle{24th Asia and South Pacific Design Automation Conference (ASPDAC '19), January 21--24, 2019, Tokyo, Japan}
\acmPrice{15.00}
\acmDOI{10.1145/3287624.3287630}
\acmISBN{978-1-4503-6007-4/19/01}
\title{Modeling Processor Idle Times in MPSoC Platforms to Enable Integrated DPM, DVFS, and Task Scheduling Subject to \\a Hard Deadline}
\titlenote{Codes and scripts for this work are available from \url{https://github.com/Amirhossein-Esmaili/Energy_Aware_Task_Scheduling_in_MPSoCs}}

\author{Amirhossein Esmaili}
\affiliation{%
  \department{Department of Electrical Engineering}
  \institution{University of Southern California}
  \city{Los Angeles}
  \state{California}
}
\email{esmailid@usc.edu}

\author{Mahdi Nazemi}
\affiliation{%
  \department{Department of Electrical Engineering}
  \institution{University of Southern California}
  \city{Los Angeles}
  \state{California}
}
\email{mnazemi@usc.edu}

\author{Massoud Pedram}
\affiliation{%
  \department{Department of Electrical Engineering}
  \institution{University of Southern California}
  \city{Los Angeles}
  \state{California}}
\email{pedram@usc.edu}

\renewcommand{\shortauthors}{}
\renewcommand{\shorttitle}{}

\begin{abstract}
 Energy efficiency is one of the most critical design criteria for modern embedded systems such as multiprocessor system-on-chips (MPSoCs). Dynamic voltage and frequency scaling (DVFS) and dynamic power management (DPM) are two major techniques for reducing energy consumption in such embedded systems. Furthermore, MPSoCs are becoming more popular for many real-time applications.
 One of the challenges of integrating DPM with DVFS and task scheduling of real-time applications on MPSoCs is the modeling of idle intervals on these platforms. In this paper, we present a novel approach for modeling idle intervals in MPSoC platforms which leads to a mixed integer linear programming (MILP) formulation integrating DPM, DVFS, and task scheduling of periodic task graphs subject to a hard deadline. We also present a heuristic approach for solving the MILP and compare its results with those obtained from solving the MILP.
\end{abstract}

%
%

\ccsdesc[500]{Computer systems organization~Embedded and cyber-physical systems}
\ccsdesc[500]{Computer systems organization~Real-time systems}
\ccsdesc[500]{Theory of computation~Scheduling algorithms}
\ccsdesc[500]{Theory of computation~Integer programming}
\ccsdesc[300]{General and reference~Design}
\ccsdesc[300]{General and reference~Performance}

\keywords{Task Scheduling, Energy Optimization, DVFS, DPM, Real-time MPSoCs}

\maketitle

\section{Introduction} \label{section.introduction}
Energy consumption is one of the most important design criteria of computing devices, ranging from portable embedded systems to servers in data centers. Furthermore, with growing demand for high performance in embedded systems, architectures such as multiprocessor system-on-chip (MPSoC) are becoming more popular for many real-time applications. 
In order to reduce energy consumption in such embedded systems, two main techniques are used, namely, dynamic voltage and frequency scaling (DVFS) and dynamic power management (DPM). In DVFS, \textcolor{black}{operating} voltage and clock frequency of processors are adjusted 
based on workload characteristics.
With DPM, processors are switched to a low power state (sleep mode) when they are not used for execution of any tasks (idle time/interval). This leads to the reduction of static power consumption. However, switching to a sleep mode has non-negligible time and energy overhead,
and it only causes energy savings when the idle time of a processor is longer than a threshold called \textit{break-even} time \cite{gerards2013optimal}.


There have been many research studies regarding reducing the energy consumption using DVFS and/or DPM. A major portion of these studies only considers DVFS for the energy optimization on single and multiprocessor platforms \cite{aydin2001determining,huang2009energy,gerards2015interplay}. Ref \cite{nakada2017energy} has focused on DPM and has proposed an energy-efficient scheduling relying on minimizing the number of processor switching and maximizing the usage of energy-efficient cores in heterogeneous platforms. Some other papers have integrated scheduling of tasks with DVFS and then, at the final phase, have applied DPM wherever it was possible \cite{srinivasan2007integer}. However, with the increase in static power portion of the total power consumption of systems \cite{huang2011applying}, both DPM and DVFS should be integrated with scheduling of tasks for the sake of energy optimization. Reference \cite{rong2006power} has combined DPM and DVFS for minimizing energy consumption of a uniprocessor platform performing periodic hard real-time tasks with precedence constraints. A major challenge of integrating DPM with the scheduling of tasks in a multiprocessor platform is formulating idle intervals and their associated energy consumption in the total energy consumption of the these platforms. The authors in \cite{chen2014energy} have developed an energy-minimization formulation for a multiprocessor system considering both DVFS and DPM and solves it via mixed integer linear programming (MILP). However, one major assumption in their formulation is that the processor assignment for the tasks to be scheduled is known in advance. Furthermore, they only consider inter-task DVFS, i.e., the frequency of the processor stays constant for the entire duration of the execution of a task. However, when there is a set of discrete frequencies available for task execution (as that is the case in \cite{chen2014energy} and also our work as we will see in Section \ref{subsection.problem_statement}), allowing intra-task DVFS and the usage of a combination of discrete frequencies for execution of tasks can result in more energy savings \cite{gerards2013optimal}.

In this paper, by proposing a method for modeling idle intervals in a multiprocessor system, we present an energy optimization MILP formulation integrating both DVFS and DPM with scheduling of real-time tasks with precedence and time constraints. By solving the MILP, for each task, we obtain the optimum processor assignment, execution start time, and the distribution of its workload among available frequencies of the processor. \textcolor{black}{To the best of our knowledge}, this is the first work that integrates both DVFS and DPM with scheduling of real-time periodic dependent tasks in a formulation that provides 
optimum values for all the aforementioned results simultaneously in a multiprocessor platform.
We also present a heuristic approach for solving the model and compare its results with those obtained from solving the MILP. 
The rest of the paper is organized as follows: Section \ref{section.Models} explains the models used for the problem formulation and presents the formal problem statement. Section \ref{section.Method} presents the proposed method and MILP formulation. Section \ref{section.Results} provides the results. Finally, Section \ref{section.Conclusion} concludes the paper and discusses \textcolor{black}{future work.}

\section{Models and Problem Definition} \label{section.Models}
\subsection{Voltage and Frequency Change Overhead}
The frequency change for modern processors takes around tens of microseconds depending on the amount and (up or down) direction of the frequency change. According to \cite{park2013accurate}, the frequency downscaling for Intel Core2 Duo E6850 processor takes approximately between 10 to 60 microseconds depending on the amount of the frequency change. In contrast, the transition to and from sleep modes of modern processors usually takes in the order of a few milliseconds. Therefore, for our modeling, we ignore the latency overhead of switching frequencies compared to that of transition to and from sleep modes of a processor. The energy overhead associated with frequency change is also small and neglected in our modeling.
\subsection{Task Model}
Tasks to be scheduled are modeled as a task graph which itself is a directed acyclic graph (DAG) represented by $G(V, E, T_d)$, in which $V$ denotes the set of tasks (we have a total of $n$ tasks), $E$ denotes data dependencies among tasks, and $T_d$ denotes the period of the task graph (i.e., tasks in the task graph are repeated after $T_d$). Each task graph should be scheduled before the arrival of the next one (i.e., $T_d$ acts as a hard deadline for scheduling of tasks). In this paper, the workload of each task is represented by the total number of processor cycles required to perform that task completely. For task $u$ ($u=1,2,...,n$), this workload is represented by $W_u$. 

\subsection{Energy Model} \label{subsection.energy_model}

For modeling the processor power consumption during executing a task with frequency $f$, similar to \cite{gerards2015interplay}, the following model would be exploited:
\begin{equation} \label{equation.power}
    P = af^{\alpha}+bf+c,
\end{equation}
in which $af^{\alpha}$ represents dynamic power portion, and $bf+c$ represents static power portion of total processor power consumption. $\alpha$ indicates the technology-dependent dynamic power exponent; usually  $\approx$ 3. $a$ is a constant that depends on the average switched capacitance and the average activity factor.
Therefore, energy consumption in one clock cycle, when executing a task with frequency $f$, is obtained via the following formulation:
\begin{equation} \label{equation.Ecycle}
    E_{cycle} = af^{\alpha-1}+b+\frac{c}{f}.
\end{equation}

For modeling the processor energy consumption during an idle time, $E_{idle}$ function is used according to the formulation presented in (\ref{equation.idleenergy}). Here, for the illustration purposes, we only use one sleep mode for switching to and waking up from, and power consumption during this sleep mode is considered to be zero (It is straightforward to extend the work to support multiple sleep modes each associated with a different non-zero power consumption):

\begin{equation} \label{equation.idleenergy}
E_{idle}(I) = \left\{ \begin{array}{cc} 
                c \times I & \hspace{5mm} 0 \leq I < T_{be} \\
                E_{sw} & \hspace{5mm} T_{be} \leq I < T_{d} \quad, \\
                0 & \hspace{5mm} I = T_{d} \\
                \end{array} \right.
\end{equation}
where $I$ represents the idle time, $c$ is the frequency-independent component of power consumption (by setting $f$ to zero in (\ref{equation.power})), and $E_{sw}$ is the switching energy overhead for both switching to the sleep mode and waking up from it. $T_{be}$ represents break-even time and is obtained as follows:
\begin{equation}
    T_{be} = \max \, (T_{sw}, \frac{E_{sw}}{c}),
\end{equation}
where $\frac{E_{sw}}{c}$ represents the minimum amount the idle time should be so that switching to the sleep mode and waking up from it causes energy savings, and $T_{sw}$ is the physical time needed for both switching to the sleep mode and waking up from it. $T_{be}$ is the maximum of these two values. Furthermore, the third term in (\ref{equation.idleenergy}) conveys the fact that if no task is assigned to a processor or equivalently $I = T_{d}$, that processor is not used for scheduling of tasks and thus does not contribute to the total energy consumption at all. Therefore, our model explores the possibility of scheduling the task graph on a subset of $K$ available processors if it results in energy savings.

\subsection{Problem Statement} \label{subsection.problem_statement}
Using the combination of DVFS and DPM, where each of these techniques can be done for each processor independently, we are looking for energy-optimized scheduling of the task graph represented by $G(V, E, T_d)$ on a platform comprising of $K$ homogeneous processors subject to a hard deadline. Each processor supports a set of $m$ distinct frequencies: $\{f_1,f_2,...,f_m\}$. We are considering a non-preemptive scheduling method. Therefore, when the execution of a task starts on each of the processors, it continues until the task completion without any interruption. Consequently, for each task, we are looking for optimum values for: processor assignment for the task, task execution start time, and distribution of the total number of required processor cycles for the complete execution of the task among $m$ available frequencies.

\section{Proposed Method} \label{section.Method}
\subsection{Constraints of the Proposed Scheduling Model} \label{subsection.constraints}
In this section, we formulate constraints of the proposed scheduling model. Duration of task $u$ ($u=1,2,...,n$) is formulated as follows:
\begin{equation} \label{equation.duration}
    Dur_u = \sum_{i=1}^{m} \frac{N_{u,i}}{f_i},
\end{equation}
where $N_{u,i}$ indicates number of processor cycles performed at $f_i$ ($i = 1, 2, ..., m$) for the execution of task $u$. Therefore:
\begin{equation}
    \sum_{i=1}^{m} N_{u,i}=W_u, \quad N_{u,i} \geq 0.
\end{equation}

According to (\ref{equation.Ecycle}) and (\ref{equation.duration}), energy consumption during the execution of task $u$ can be formulated as follows:
\begin{equation} \label{equation.task_energy}
    E_{task}(u) = \sum_{i=1}^{m}(N_{u,i}.(af_i^{\alpha-1}+b+\frac{c}{f_i})).
\end{equation}

To ensure each task finishes its execution before $T_d$, for $u = 1, 2, ..., n,$ we have:
\begin{equation}
    S_u + Dur_u \leq T_d, \quad S_u \geq 0,
\end{equation}
where $S_u$ represents start time of the execution of task $u$. Furthermore, the precedence constraint is formulated as follows:
\begin{equation}
    S_u + Dur_u \leq S_v, \quad \forall e(u,v) \in E.
\end{equation}
Here, we do not consider any inter-task communication cost associated with $e(u,v)$ for sending output data of task $u$ to input data of task $v$ (The model can be easily extended to incorporate this cost). 

For processor assignment for task $u$ to processor $k, k = 1, 2, ..., K,$ we introduce the decision variable of $P_{k,u}$ which is defined as follows:
\begin{equation}
    P_{k,u} = \left\{ \begin{array}{cc} 
                1 & \hspace{5mm} \textrm{if task u is assigned to processor k} \\
                0 & \hspace{5mm} \textrm{otherwise} \\
                \end{array} \right..
\end{equation}
Therefore, we have the following constraint:
\begin{equation}
    \sum_{k=1}^{K} P_{k,u}=1, \quad \textrm{for } u = 1, 2, ..., n.  
\end{equation}
One other important constraint that needs to be satisfied is that the execution of tasks assigned to the same processor shall not overlap each other (non-preemptive scheduling). For this, we define an auxiliary decision variable called $O_{k,u,v}$ representing ordering of tasks. For $k = 1, 2, ..., K$; $u = 1, 2, ..., n$; $v = 1, 2, ..., n, v \neq u;$ we define:
\begin{equation} \label{equation.O}
    O_{k,u,v} = \left\{ \begin{array}{ll} 
                1 & \hspace{1mm} \textrm{if task u is scheduled immediately} \\
                 & \hspace{1mm} \textrm{ before task v on processor k} \\
                 & \hspace{1mm}  \\
                0 & \hspace{1mm} \textrm{otherwise} \\
                \end{array} \right..
\end{equation}
In addition, if task $v$ is the first task assigned to processor $k$, we define $O_{k,0,v}$ to be 1 (and is 0 otherwise). On the other hand, if task $u$ is the last task assigned to processor $k$, we define $O_{k,u,n+1}$ to be 1 (and is 0 otherwise).
Furthermore, if there is no task assigned to processor $k$, we define $O_{k,0,n+1}$ to be 1 (and is 0 otherwise). 
Accordingly, using \eqref{equation.O} and the definitions provided for $O_{k,0,v}$, $O_{k,u,n+1}$ and $O_{k,0,n+1}$, we have the following constraints for $k = 1, 2, ..., K$:
\begin{equation} \label{equation.next}
     \sum_{\substack{v=1 \\ v\neq u}}^{n+1} O_{k,u,v} = P_{k,u}, \quad \textrm{for } u = 0, 1, ...,n 
\end{equation}
\begin{equation} \label{equation.prev}
    \sum_{\substack{u=0 \\ u\neq v}}^{n} O_{k,u,v} = P_{k,v}, \quad \textrm{for } v = 1, 2, ..., n+1.
\end{equation}
According to (\ref{equation.next}), if task $u$ is assigned to processor $k$ ($P_{k,u}=1$), either there is one and only one task scheduled immediately after task $u$ on processor $k$ or task u is the last task assigned to processor $k$. Similarly, according to (\ref{equation.prev}), if task $v$ is assigned to processor $k$ ($P_{k,v}=1$), either there is one and only one task scheduled immediately before task $v$ on processor $k$ or task $v$ is the first task assigned to processor $k$. In both (\ref{equation.next}) and (\ref{equation.prev}), $P_{k,0}$ and $P_{k,n+1}$ are defined as 1 for all $k = 1, 2, ..., K$. 

For non-preemptive scheduling we should have:
\begin{equation}
\begin{split}
    \sum_{k=1}^K{\sum_{u=1}^{n}{\left(\left(S_u+Dur_u\right).O_{k,u,v}\right)}} \leq S_v,  \\ 
    \textrm{for } v = 1, 2, ..., n, v \neq u,
\end{split}
\end{equation}
which can be formulated as the following linear constraint:
\begin{equation} \label{equation.nonpreemptive}
\begin{split}
    S_u + Dur_u - (1 - O_{k,u,v}) \times T_d \leq S_v, \\
    \textrm{for } u = 1, 2, ..., n, \\
    \textrm{for } v = 1, 2, ..., n, v \neq u, \\
    \textrm{for } k = 1, 2, ..., K.
\end{split}
\end{equation}

\subsection{Modeling Idle Intervals}
Using $O_{k,u,v}$ variables introduced in Section \ref{subsection.constraints}, we can conveniently model idle intervals in an MPSoC platform. Specifically, for each task $v$ ($v = 1, 2, ..., n$), we formulate the amount of the idle time before servicing task $v$ on the processor to which task $v$ is assigned. When task $v$ is not the first task scheduled on the processor to which it is assigned, the idle time before servicing task $v$ can be written as follows:
\begin{align} \label{equation.I}
  I_v&=(1-\sum_{k=1}^{K}O_{k,0,v})  \nonumber \\
  &\times(S_v-\sum_{k=1}^{K}\sum_{\substack{u=1 \\ u\neq v}}^{n}((S_u+Dur_u).O_{k,u,v})).
\end{align}
If task $v$ is the first task scheduled on any of $K$ processors, the first term in multiplication in (\ref{equation.I}) causes $I_v$ to be zero. In that case, idle time before servicing task $v$ on the processor $k$ to which the task is assigned is obtained using the following:
\begin{align} \label{equation.I'}
     I'_k &= (T_d-\sum_{u=1}^{n}((S_u+Dur_u).O_{k,u,n+1})) \nonumber \\
     &+ \sum_{v=1}^{n}(O_{k,0,v} \times S_v).
\end{align}
In (\ref{equation.I'}), the second term in summation represents the idle time on processor $k$ before servicing its first assigned task in the current period. On the other hand, The first term in the summation in (\ref{equation.I'}) represents the idle time on that processor after servicing its last assigned task in the previous period. This interval should also be taken into account when calculating the amount of idle time before servicing first task scheduled on processor $k$. 
If there is no task assigned to processor $k$ at all, (\ref{equation.I'}) would give the value of $T_d$ for $I'_{k}$.
\subsection{Objective Function} \label{section.objective_function}
Subject to constraints formulated so far, we are trying to minimize the following objective function which represents the total energy consumption:
\begin{equation} \label{equation.objective}
    \sum_{u=1}^{n}E_{task}(u) + \sum_{v=1}^{n}E_{idle}(I_v) + \sum_{k=1}^{K}E_{idle}(I'_k).
\end{equation}
The objective function of (\ref{equation.objective}), alongside the formulated constraints, forms a mixed integer programming over the positive real variables of $S_u$ and $N_{u,i}$; and the Boolean decision variables of $P_{k,u}$ and $O_{k,u,v}$. The number of these variables in our problem are $n$, $nm$, $nk$, and $(n+1)^2k-nk$, respectively. 

However, due to formulations presented for idle time intervals in  (\ref{equation.I}) and  (\ref{equation.I'}), and the concave piece-wise behavior of $E_{idle}$ function in (\ref{equation.idleenergy}) in a minimization problem, it is a non-linear non-convex programming ($E_{task(u)}$ in (\ref{equation.objective}) is linear with respect to positive real variables of $N_{u,i}$ and this term does not contribute to the nonlinearity of the problem). 

For linearizing (\ref{equation.I}) and (\ref{equation.I'}), we use the lemma mentioned in \cite{chen2014energy}, where this lemma is stated as follows: 
\textit{Given constants $s_1$ and $s_2$, if $P_1$ and $P_2$ are two constraint spaces where $P_1$ is $\{[t,b,x] \, | \, t=bx, \,-s1 \leq x \leq s2, \, b \in \{0,1\}\}$, and $P_2$ is $\{[t,b,x] \, | \, -bs_1 \leq t \leq bs_2, \,t+bs_1-x-s_1 \leq 0, \,t-bs_2-x+s_2 \geq 0, \, b \in \{0,1\}\}$, then, $P_1$ and $P_2$ are equivalent.} \textcolor{black}{Proof of this lemma is given in \cite{chen2014energy}.}
With this lemma, we can substitute multiplication of a Boolean decision variable and a bounded real variable, with a newly introduced bounded real variable and \textcolor{black}{three} added linear constraints indicated in $P2$). Using this lemma multiple times, we can reach linear representations for idle time interval formulations in (\ref{equation.I}) and (\ref{equation.I'}) at the end.

Furthermore, $E_{idle}(I_v)$ in (\ref{equation.objective}) can be written as follows:
\begin{equation} \label{equation.idleenergy_v}
    E_{idle}(I_v) = S_v.(E_{sw}) + (1-S_v).(c \times I_v),
\end{equation}
where $S_v$ is a Boolean decision variable which is 1 when $T_{be} \leq I_v < T_d$ and is 0 otherwise ($I < T_{be}$). Therefore, this decision variable represents switching and whether we put the processor during $I_v$ in the sleep mode or not. Since $I_v$ represents the amount of idle time before servicing task $v$ on the processor to which task $v$ is assigned when task $v$ is not the first task on that processor, $I_v$ can never be $T_d$. Therefore, we do not need to formulate the third term of (\ref{equation.idleenergy}) in (\ref{equation.idleenergy_v}). Corresponding constraint for $S_v$ ($v = 1, 2, ..., n$), is written as follows:
\begin{equation}
    \frac{I_v - T_{be}}{T_d} \leq S_v \leq \frac{I_v}{T_{be}}, \quad S_v \in \{0,1\}.
\end{equation}

For $E_{idle}(I'_k)$, a similar formulation like  (\ref{equation.idleenergy_v}) can be used except that we need another Boolean decision variable called $U_k$ which represents whether we assign any task to processor $k$ or not. When $U_k$ is 0, it means processor $k$ is not used at all for scheduling the task graph and thus does not contribute to the energy consumption in (\ref{equation.objective}). Therefore, $U_k$ is 1 when we assign one or more tasks to processor $k$ ($I'_k < T_d$) and is 0 otherwise ($I'_k = T_d$, or equivalently: $T_d \leq I'_k \leq T_d$). Accordingly, $E_{idle}(I'_k)$ in (\ref{equation.objective}) can be written as follows:
\begin{equation} \label{equation.idleenergy_k}
    E_{idle}(I'_k) = U_k.[S'_k.(E_{sw}) + (1-S'_k).(c \times I'_k)],
\end{equation}
where $S'_k$ represents whether we switch the processor during $I'_k$ to the sleep mode or not (similar to $S_v$). The usage of $U_k$ in (\ref{equation.idleenergy_k}) allows formulating the third term of (\ref{equation.idleenergy}). Corresponding constraints for $S'_k$ and $U_k$ ($k = 1, 2, ..., K$) are written as follows:
\begin{equation}
    \frac{I'_k - T_{be}}{T_d} \leq S'_k \leq \frac{I'_k}{T_{be}}, \quad S'_k \in \{0,1\},
\end{equation}
\begin{equation}
    \frac{I'_k - T_d}{Td} \leq U_k \leq \frac{I'_k}{T_d}, \quad U_k \in \{0,1\}.
\end{equation}

In order to linearize
(\ref{equation.idleenergy_v}) and (\ref{equation.idleenergy_k}), we again use the aforementioned lemma. However, for (\ref{equation.idleenergy_k}), where we have a multiplication of two Boolean decision variables, we also need the following lemma:
\textit{If $P_1$ and $P_2$ are two constraint spaces where $P_1$ is $\{[z,x,y] \, | \, z=xy, \, x \in \{0,1\}, \, y \in \{0,1\} \}$, and $P_2$ is $\{[z,x,y] \, | \, z \leq x, \, z \leq y, \, x+y-z \leq 1, \, x \in \{0,1\}, \, y \in \{0,1\}\}$, then, $P_1$ and $P_2$ are equivalent.}
Using these lemmas and methods for linearizing the objective function of (\ref{equation.objective}), the energy-optimized scheduling problem expressed in \textcolor{black}{Section \ref{subsection.problem_statement}} is modeled as an MILP formulation. 

\section{Results} \label{section.Results}
\subsection{Experiment Setup}
In order to solve the formulated MILP, we use IBM ILOG CPLEX Optimization Studio \cite{cplex}. The platform on which simulations are performed is a computer with a 3.2 GHz Intel Core i7-8700 Processor and 16 GB RAM.
\textcolor{black}{Using \cite{chen2014energy}} for obtaining energy model parameters, the frequency-independent component of processor power consumption, which is represented by $c$ in (\ref{equation.power}), is obtained as $276\,mW$. Each processor can operate independently of other processors at either $f_1=1.01\,GHz$, $f_2=1.26\,GHz$, $f_3=1.53\,GHz$, $f_4=1.81\,GHz$, $f_5=2.1\,GHz$. For these frequencies, frequency-dependent component of processor power consumption, which is represented by $af^\alpha+bf$ in (\ref{equation.power}), is $430.9\,mW$, $556.8\,mW$, $710.7\,mW$, $896.5\,mW$, and $1118.2\,mW$, respectively. Using curve fitting, we obtain $a=23.8729$, $b=401.6654$, and $\alpha=3.2941$ in (\ref{equation.power}).  
$E_{sw}$ and $T_{sw}$ are set as $385\,\mu J$ and \textcolor{black}{$5\,ms$}. Here, We consider a architecture with 4 processors. Simulations are performed on \textcolor{black}{8} task graphs randomly generated using TGFF \cite{dick1998tgff}, which is a randomized task graph generator widely used in the literature to evaluate the performance of scheduling algorithms.  
Detailed information for each task graph is presented in Table \ref{table.results}. For studied task graphs, the average workload of each task is set to $2 \times 10 ^ 6$ cycles (around $1\,ms$ execution time under maximum frequency). The maximum in-degree and out-degree for each node is set to 2 and 3, respectively. The number of tasks in studied random task graphs ranges from \textcolor{black}{7 to 28}.

To evaluate the advantage of our modeling of idle intervals in multiprocessor systems, we consider two cases: 
1) A baseline case which uses only the first term of (\ref{equation.objective}) as the objective function alongside with constraints of (\ref{equation.duration}) to (\ref{equation.prev}), and (\ref{equation.nonpreemptive}). In other words, in this baseline case, we do not use any idle time-related terms in the objective function or constraints. 2) Using (19) as the objective function alongside all constraints and linearization techniques mentioned in Section \ref{section.Method} (this case is our proposed method).
In the baseline case, switching to the sleep mode during an idle time is done, if possible, after the scheduling is finished (i.e., DPM is not integrated with DVFS and scheduling in the baseline case). Therefore, the baseline case is an \underline{i}ntegrated \underline{S}cheduling and \underline{C}lock-and-voltage scaling followed by mode \underline{T}ransition algorithm (iSC+T). The second case, which is our proposed method, is referred to as an \underline{i}ntegrated \underline{S}cheduling, \underline{C}lock-and-voltage scaling, and mode \underline{T}ransition algorithm (iSCT).

%

\begin{table} [tb]
\centering
\captionsetup{justification=centering}
\renewcommand{\arraystretch}{1.05}
\caption{\textcolor{black}{\footnotesize Task Graphs Characteristics and Corresponding Energy Consumption Values Obtained from iSCT versus iSC+T}}
\label{table.results}
\begin{center}
\scalebox{0.60}{
    \begin{tabular}{|c|c|c|c|c|c|c|}
        \hline
        Task&No. of&Total Workload of &Td&\multicolumn{2}{c|}{Total Energy}&iSCT versus iSC+T\\
        Graph&Tasks &Tasks in Processor& (ms)& \multicolumn{2}{c|}{Consumption (mJ)} & Energy Saving \\
        \cline{5-6}
             &         & Cycles ($\times 10^6$)      &     &    iSC+T  &    iSCT  &  (\%)    \\
        
        \hline
        TGFF1&7&15.89&8&12.67&10.45&17.52\\
        \hline
        TGFF2&11&18.69&12&13.60&12.06&11.32\\
        \hline
        TGFF3&14&34.39&10&25.99&22.70&12.66\\
        \hline
        TGFF4&15&31.89&12&28.08&21.00&25.21\\
        \hline
        TGFF5&16&34.88&12&27.35&23.02&15.83\\
        \hline
        TGFF6&18&33.46&14&25.38&21.98&13.40\\
        \hline
        TGFF7&22&44.94&22&33.74&29.39&12.89\\
        \hline
        TGFF8&28&56.81&18&42.95&37.00&13.85\\
        \hline        
    \end{tabular}
}
\end{center}
\end{table}

\subsection{Effect of Modeling Idle Intervals}
According to Table \ref{table.results}, including the energy consumption of modeled idle intervals in the objective function causes an average energy saving of \textcolor{black}{15.34\% (up to 25.21\%)} for iSCT versus iSC+T.
To better observe the contribution of modeling idle intervals in an MPSoC platform, for each scheduled task graph, the total number of idle intervals on all processors, and total time of these idle intervals are shown in Table \ref{table.idle} for both iSCT and iSC+T. Furthermore, for each scheduled task graph, the number of used processors for the scheduling of that task graph, out of maximum 4 processors, is also presented in Table \ref{table.idle} for both iSCT and iSC+T.  

According to Table \ref{table.idle}, while for all scheduled task graphs, the total time of idle intervals are higher or the same for iSCT compared to iSC+T, the number of idle intervals for iSCT are notably fewer than the number of idle intervals for iSC+T \textcolor{black}{(on average fewer than half)}. \textcolor{black}{Therefore, by including the energy consumption of modeled idle intervals in the objective function of (\ref{equation.objective}), instead of having a number of distributed short idle intervals, we will have fewer merged longer idle intervals.}
This results in more opportunities for switching the processors to the sleep mode during idle intervals and thus more energy savings (as indicated in Table \ref{table.results}). In fact, for task graphs studied in this paper, the percentage of idle intervals that are longer than $T_{be}$, and thus we can switch the processor to the sleep mode, are \textcolor{black}{91.67 \% and 32.31 \%} for iSCT and iSC+T, respectively.   

On the other hand, as indicated in Table \ref{table.idle}, iSCT explores the possibility of the usage of a subset of 4 processors if it results in energy savings. \textcolor{black}{For discussed task graphs in this paper, iSCT always uses fewer than 4 processors for the scheduling.} The unused processors do not contribute to the energy consumption at all (we do not need to switch them to the sleep mode and wake them up in every time period). 
This can be helpful in terms of energy efficiency, \textcolor{black}{particularly when the energy overhead of switching processors is relatively high.} Reference \cite{chen2014energy} cannot take advantage of this since the processor assignment for tasks to be scheduled is assumed to be known in advance and it is not integrated in their MILP formulation.

\textcolor{black}{On the platform we performed simulations,} iSCT and iSC+T approaches on average generated results for studied task graphs in less than \textcolor{black}{69 and 1} minutes, respectively. Since we are considering scheduling of a periodic task graph, these simulations are done offline only once for one period of the task graph. The obtained scheduling can be programmed to a MPSoC for real-time scheduling of each arriving period of the task graph.
\begin{table}[tb]
    \centering
    \captionsetup{justification=centering}
    \renewcommand{\arraystretch}{1.05}
    \caption{\textcolor{black}{\footnotesize Idle Intervals Characteristics and No. of Used Processors for iSCT versus iSC+T}}
    \label{table.idle}
\scalebox{0.60}{
    \begin{tabular}{|c|c|c|c|c|c|c|}
         \hline
         Task  & \multicolumn{2}{c|}{No. of Idle Intervals}&\multicolumn{2}{c|}{Total Idle time (ms) }&\multicolumn{2}{c|}{No. of Used Processors} \\
         \cline{2-7}
         Graph & iSC+T & iSCT & iSC+T & iSCT & iSC+T & iSCT\\             
         \hline
         TGFF1& 5 & 3 & 21.62 & 24.00&3&1\\
         \hline
         TGFF2& 4 & 3 & 35.79 & 36.00&4&1\\
         \hline
         TGFF3& 6 & 3 & 17.53 & 21.22&4&2\\
         \hline
         TGFF4& 10 & 3 & 27.16 & 29.00&4&2\\
         \hline
         TGFF5& 7 & 3 & 25.21 & 29.00&4&2\\
         \hline
         TGFF6& 8 & 3 & 34.13 & 34.13&4&2\\
         \hline
         TGFF7& 10 & 3 & 58.63 & 58.64&4&2\\
         \hline
         TGFF8& 10 & 3 & 34.87 & 36.96&4&2\\
         \hline         
    \end{tabular}
}
\end{table}
\subsection{A Heuristic Approach to Solve the Model}
Here, we propose a two-stage heuristic algorithm to solve the formulated model: 1) We first determine and fix the values of $O_{k,u,v}$ and $P_{k,u}$ variables using a polynomial-time list scheduling algorithm. 2) Using fixed $O_{k,u,v}$ and $P_{k,u}$ values, the number of variables in the original MILP problem reduces significantly. Also, (\ref{equation.I}) and (\ref{equation.I'}) will become linear formulations in the first place, and we do not anymore need to use \textcolor{black}{the first lemma presented in Section \ref{section.objective_function}} multiple times to linearize them. This further reduces the number of variables of the MILP problem considerably (\textcolor{black}{since each time using of this lemma adds one set of real variables, plus three sets of constraints}). Then, the new formulated problem with considerably fewer number of variables, which is still an MILP due to the usage of (\ref{equation.idleenergy_v}) and (\ref{equation.idleenergy_k}) in the objective function of (\ref{equation.objective}), will be solved to obtain values of $S_u$ and $N_{u,i}$. 

For the first stage, we use a variant of heterogeneous earliest finish time (HEFT) algorithm \cite{topcuoglu2002performance}. While this algorithm aims for heterogeneous platforms, it can be applied to a homogeneous platform similar to our paper as well. Basically, in this algorithm, tasks are ordered according to their \textit{upward rank}, which is defined recursively for each task 
as follows:
\begin{equation}
    rank_{up}(u) = Dur^{*}_{u} + \max_{v \in succ(u)}(rank_{up}(v)),   
\end{equation}
where here, $Dur^{*}_{u}$ is the duration of the task $u$ when all of its workload is executed using maximum available frequency, and $succ(u)$ is the set of immediate successors of task $u$. Ranks of the tasks are computed recursively starting from \textit{exit tasks} of the task graph (exit tasks are the ones with out-degree of zero). The upward rank of exit tasks are equal to their corresponding $Dur^{*}$ values. Basically, $rank_{up}(u)$ indicates the length of critical path from task $u$ to exit tasks, including $Dur^{*}_{u}$ itself. 

After the calculation of ranks for all the tasks, a task list is generated by sorting the tasks in the decreasing order of their ranks. Tie-breaking is done randomly. Then, tasks are scheduled on processors based on the order of the task list. Each task can only be scheduled after a time called \textit{ready\_time} of that task, which indicates the time that the execution of all immediate predecessors of that task has completed. For each task, we look for the first idle interval on each processor after the task ready\_time, with the amount of at least $Dur^{*}$ of that task, and assign the task to the processor which gives us the earliest finish time. \textcolor{black}{Since} the task list sorted by the decreasing order of ranks gives a topological sorting of the DAG \cite{topcuoglu2002performance}, when we choose a task for scheduling, its predecessors have already been scheduled. The time complexity of HEFT algorithm is $O(|E| \times K)$ where $|E|$ denotes the number of edges of the DAG and $K$ denotes the number of processors \cite{topcuoglu2002performance}.

Using HEFT in the first stage of our heuristic approach, we determine the processor assignment for each task ($P_{k,u}$), and ordering of tasks on each processor ($O_{k,u,v}$). Note that obtained start times for tasks after the first stage just show relative ordering of tasks on each processor. Also, we only used maximum frequency in the first stage. Next, in the second stage, we solve the newly derived MILP, which has been obtained after fixing $O_{k,u,v}$ and $P_{k,u}$ values in the first stage, and has considerably fewer number of variables compared to the original MILP. This gives us the values for $S_{u}$ and $N_{u,i}$ variables. On average, on the platform we performed simulations, solving the newly derived MILP provided results for studied task graphs in less than \textcolor{black}{2 seconds}, which is considerably less than the simulation time of solving the original MILP.

\begin{figure}[!t]
\centering
\captionsetup{justification=centering}
\includegraphics[width=0.5\textwidth]{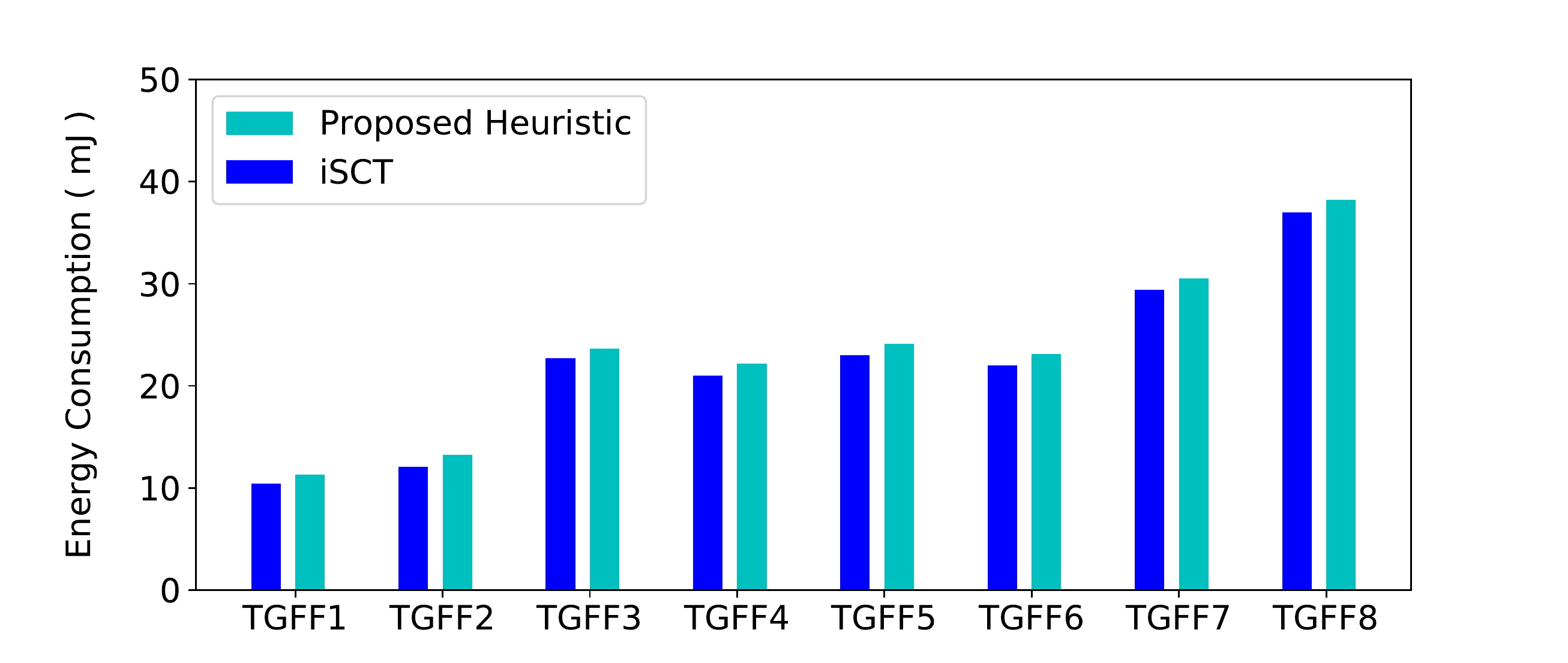}
\caption{\textcolor{black}{ \footnotesize Energy Consumption obtained from the proposed heuristic approach and iSCT for different task graphs}} \label{fig.heft}
\end{figure}

Fig. \ref{fig.heft} shows a comparison between the energy consumption obtained from iSCT,
and the energy consumption obtained from solving the problem using the proposed heuristic approach. According to Fig. \ref{fig.heft}, the heuristic method provides close estimates compared to the optimum solution. The values of energy consumption obtained from the heuristic approach are on average \textcolor{black}{5.66\%} higher than the optimum solution.      

\balance
\section{Conclusions and Future Work} \label{section.Conclusion}

\textcolor{black}{In this paper}, by proposing a method for modeling idle intervals in multiprocessor systems, we presented an energy optimization MILP formulation integrating both DVFS and DPM with scheduling of real-time tasks with precedence and time constraints. By solving the MILP, for each task, we obtain the optimum processor assignment, execution start time, and the distribution of its workload among available frequencies of the processor. Results show the effectiveness of our modeling of idle intervals in MPSoCs in terms of energy efficiency. We also presented a heuristic approach for solving the MILP which provided close results compared to optimum results. It is worth mentioning that although our proposed model focuses on MPSoCs, it can also be applicable to servers in data centers by using proper energy model parameters of those platforms.

For future work, workload of tasks can be investigated to represent more than just the processor cycle count; e.g., the memory requirement, or the possibility of executing the entire or part of a task on GPUs can be modeled and investigated. Also, obtaining a variant of the proposed model for heterogeneous processors could be another potential future direction.

\bibliographystyle{unsrt}

\end{document}